\documentclass[twocolumn,aps,pra,showpacs,preprintnumbers,letterpaper,amssymb,amssymb,floatfix,superscriptaddress]{revtex4-2}

\usepackage[T1]{fontenc}
\usepackage{natbib}
\usepackage{graphicx}
\usepackage{bm}
\usepackage{color}
\usepackage{amsmath}
\usepackage{xspace}
\usepackage{subfigure}
\usepackage{dcolumn}
\usepackage{textcomp}
\usepackage{longtable}
\usepackage{url}
\usepackage[colorlinks,urlcolor=blue,citecolor=blue,linkcolor=blue]{hyperref}
\usepackage{bm}
\usepackage{appendix}
\usepackage{hyperref}
\usepackage{soul}
\usepackage[normalem]{ulem}
\usepackage{physics}
\usepackage{tabularx}
\usepackage{array}
\usepackage{svg}

\definecolor{bggreen}{RGB}{185,230,70}
\definecolor{myblue}{RGB}{0, 40, 140}

\begin{document}

\title{Acousto-optic lens for 3D shuttling of atoms in a neutral atom quantum computer}
\author{Z. \surname{Guo}}
\email{z.guo1@tue.nl}
\author{R.A.H. \surname{van Herk}}
\author{E.J.D. \surname{Vredenbregt}}
\author{S.J.J.M.F. \surname{Kokkelmans}}
\affiliation{Department of Applied Physics and Science Education, Eindhoven University of Technology, P. O. Box 513, 5600 MB Eindhoven, The Netherlands}
\affiliation{Eindhoven Hendrik Casimir Institute, Eindhoven University of Technology, P. O. Box 513, 5600 MB Eindhoven, The Netherlands}

\date{\today}

\begin{abstract} 
We present a novel acousto-optic lens (AOL) design for neutral atom quantum computing. This approach enhances atom rearrangement in optical tweezer arrays and addresses the speed limitations imposed by the cylindrical lensing effect of acousto-optic deflectors (AODs). By combining a double-pass AOD configuration for dynamic focal tuning with a standard pair of crossed AODs for transverse beam steering, our design enables the generation of arbitrary focal point trajectories. This configuration enables shuttling of atoms in 3D space, thereby helping to realise fully connected two-qubit gates and mid-circuit measurements. We detail the optical implementation, characterize its performance, and discuss its applications in scalable quantum computing architectures.
\end{abstract}

\maketitle

\section{Introduction}

\begin{figure}[htb]
\centering
\includegraphics[width=\columnwidth]{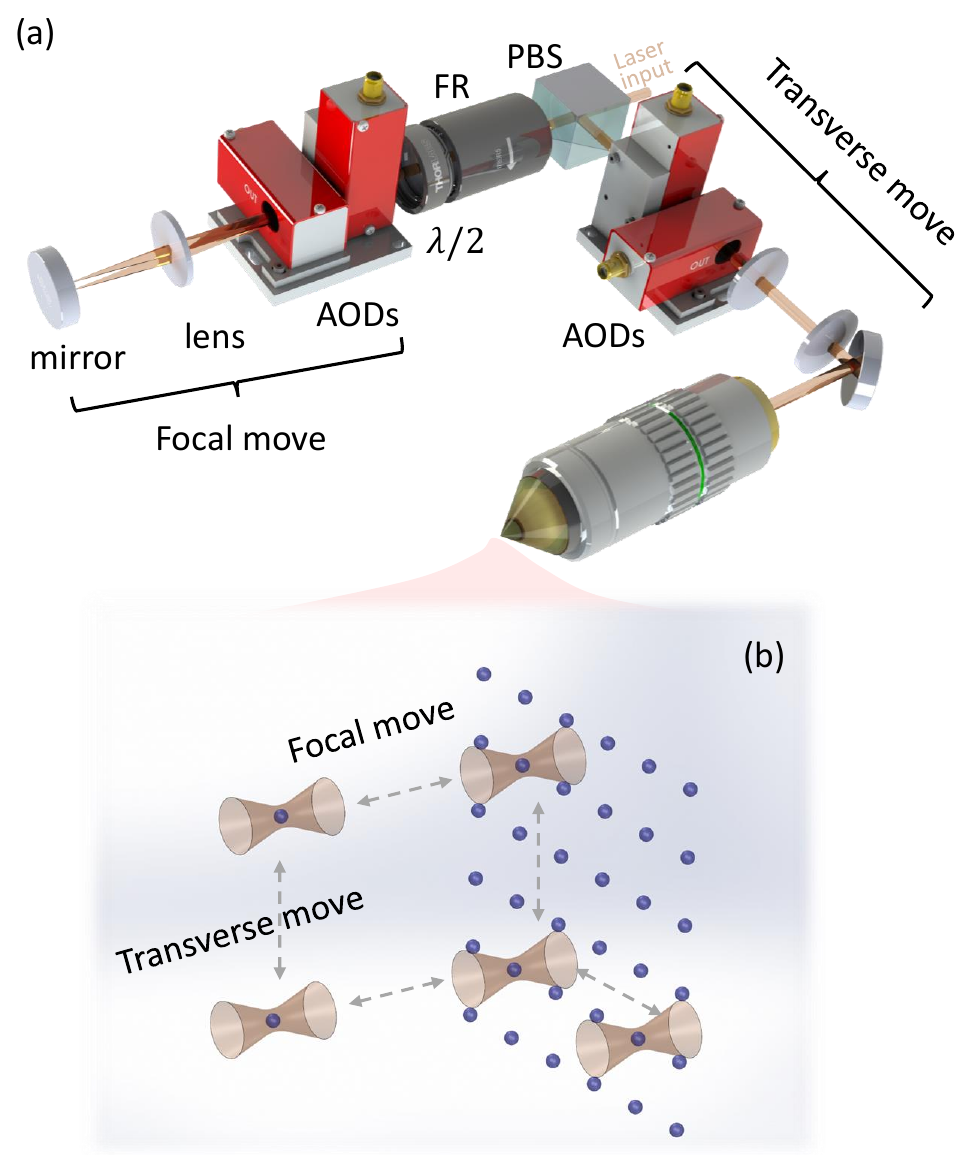}
\caption{Design of a variant of an acousto-optical lens (AOL). (a) The laser beam enters from the right side into the PBS and passes through the Faraday rotator and a half-wave plate, which together function as an optical circulator. The use of this configuration is primarily to ensure polarization matching with the AOD. After double-passing through the crossed AOD, the laser beam proceeds through a normal crossed AOD for transverse movement and is finally focused by the microscope. (b) Illustrates both the focal and transverse move of atoms enabled by the AOL. In a conventional crossed-AOD setup, movable tweezers can only shuttle atoms within a single focal plane, restricting the achievable spacing between traps. By contrast, the AOL allows atoms to be moved out of the focal plane, enabling full 3D shuttling.}
\label{Fig:AOL}
\end{figure}

The recent progress of neutral atom quantum computing \cite{bernien_probing_2017} has seen pivotal milestones such as the establishment of high-fidelity gates \cite{evered_high-fidelity_2023,scholl_erasure_2023}, the development of large-scale arrays \cite{manetsch_tweezer_2024, wurtz_aquila_2023}, mid-circuit measurements \cite{singh_mid-circuit_2023-1,singh_dual-element_2022,ma_high-fidelity_2023, muniz_repeated_2025}, and significant progress in quantum error correction \cite{bluvstein_logical_2023, sales_rodriguez_experimental_2025}. These advances emphasize the critical need for system scalability, qubit addressability, and parallelisability. A key component in these advancements is atom shuttling, which involves moving atoms using an optical tweezer controlled by 2D acousto-optic deflectors (AODs). AOD-based control has proven particularly effective for dynamically rearranging atoms in optical tweezer arrays~\cite{barredo_atom-by-atom_2016} and for implementing two-qubit entangling operations~\cite{bluvstein_quantum_2022}. These capabilities complement alternative approaches, such as spatial light modulators (SLMs), which have also been used in atom transport~\cite{lee_three-dimensional_2016, knottnerus_parallel_2025}.

Atom transport \cite{barredo_atom-by-atom_2016} using crossed AODs is accomplished by adjusting the radio frequencies (RF) driving two cross-aligned AODs, directing a tweezer to different points in the focal plane of the microscope. For dense tweezer arrays with separations below 5~$\mu$m, established protocols generally move atoms directly from site to site while avoiding occupied neighboring sites \cite{barredo_synthetic_2018-1}. Consequently, atom rearrangement algorithms are designed to route tweezers around filled sites, requiring sophisticated optimisation strategies \cite{mamee_heuristic_2021,sheng_defect-free_2022,lee_defect-free_2017}. Atom movement is also used for the implementation of any-to-any entanglement gates \cite{bluvstein_logical_2023} and mid-circuit measurements \cite{singh_mid-circuit_2023-1}. These techniques require coherent transport and therefore require large tweezer separation \cite{bluvstein_quantum_2022,chiu_continuous_2025}, which either requires an additional approach to expand the tweezer array during a circuit, or a decrease in the total number of tweezer sites. Having a large tweezer separation also results in longer transport distances and potentially a lower atom survival chance \cite{evered_high-fidelity_2023,bluvstein_logical_2023}. Previous work set separate regions for atom loading and imaging, preventing the scattering of photons onto other atoms \cite{norcia_mid-circuit_2023}, leading to even longer distance moves. The lensing effect of the AOD \cite{vanderlugt_optical_2005} deforms the tweezer trap potential when the RF frequency ramps rapidly, setting a speed limit for the moving tweezer.

To overcome the constraints of planar atom shuttling, we introduce a fast, three‑dimensional transport scheme that employs a double‑pass configuration of an additional set of crossed AODs, inspired by two‑photon excitation fluorescence microscopy (TPEF) \cite{reddy_fast_2005,kaplan_acousto-optic_2001}. In TPEF, four AODs are arranged as two crossed pairs, each pair simultaneously acting as a tunable cylindrical lens and a transverse beam scanner. Transverse displacements arise from straightforward shifts of the RF drive frequency, while longitudinal refocusing is achieved by applying a linear frequency chirp, which creates a position‑dependent diffraction angle in the crystal and thus an effective lens. This inertia‑free acousto‑optic steering architecture enables volumetric scan rates above 50 kHz \cite{katona_fast_2012,duemani_reddy_three-dimensional_2008}. However, adapting the concept to neutral atom tweezer experiments requires AOD crystals engineered for a broad range of incident angles \cite{kirkby_compact_2010} and an RF synthesis system capable of generating intricate coupled RF signal for transverse and longitudinal movement \cite{reddy_fast_2005,kirkby_compact_2010}.

Our acousto-optic lens (AOL) redesign mitigates these issues by decoupling transverse steering from longitudinal refocusing (Fig. \ref{Fig:AOL}(a)). Transverse motion is provided by a crossed AOD pair, while longitudinal refocusing is realized with a double‑pass through an additional pair of crossed AODs. The double‑pass geometry eliminates the change in output beam direction during focus shifts and self‑corrects the AOD phase matching (Fig. \ref{Fig:lensing effect}(c)), removing the need for specially engineered wide‑angle AODs \cite{kirkby_compact_2010}. The trade‑off is an extra pass through the crossed AOD, leading to an acceptable power loss of around 30\%. Using only off‑the‑shelf AODs and standard optics, we demonstrate this AOL variant, which can facilitate fast three‑dimensional atom shuttling.

In this work, we optically test our new AOL and demonstrate the ability to achieve programmable focus position movement. We propose several use cases in neutral atom quantum computing, such as facilitating atom rearrangement in tweezer arrays, speeding up the coherent transport for a fully connected two-qubit gate and mid-circuit measurements. We also investigate the lensing effect that arises in the standard crossed‑AOD configuration for transverse atom shuttling under rapid RF ramps, evaluate how strongly it deforms the traps and how it induces the motional excitation of the trapped atom. Our new setup can compensate for this additional AOD lensing, thereby improving atom transport.

The structure of the remaining paper is as follows: Section \ref{sec: review} reviews the operating principles of AOLs and summarizes AOL architectures developed in the TPEF community. Section \ref{sec: AOL design} introduces the AOL variants we propose. Section \ref{sec: characterization} presents the optical characterization of our system. Section \ref{sec: applications} discusses several applications in neutral atom quantum computing.

\section{Short review of acousto-optic lenses}\label{sec: review}

\begin{figure*}[htbp]
\centering
\includegraphics[width=0.7\linewidth]{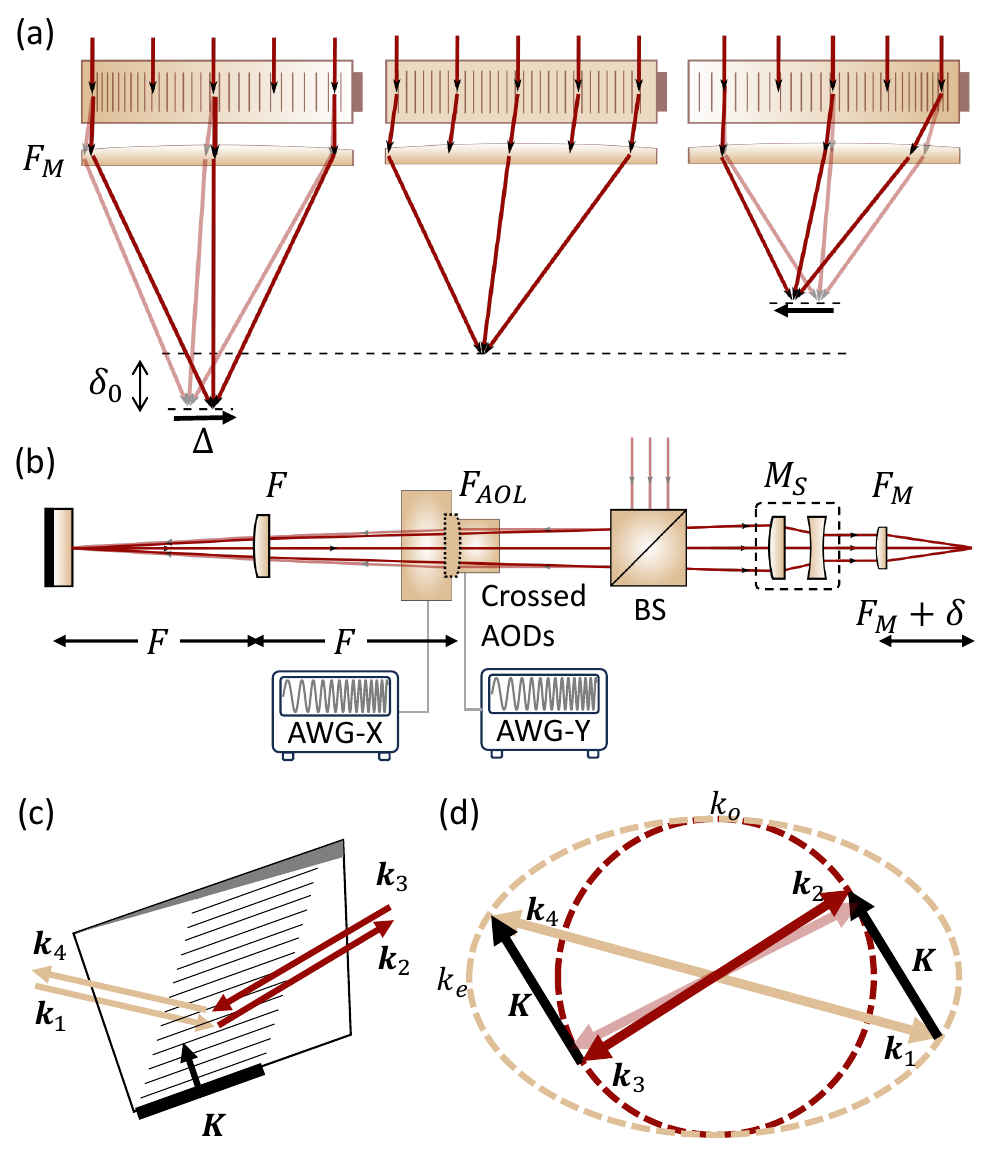}
\caption{Lensing effect of an AOD and double-pass configuration for focus adjustment. (a) Laser beam passing through the AOD crystal is focused by an additional lens $F_M$. The center graph shows injection with a constant RF frequency, which only bends the output beam angle without affecting the focal position. The left (right) graph shows up (down) chirping of the RF frequency, resulting in focus extension (shortening) $\delta_0$. Note that there is always a time-varying transverse movement $\Delta$ due to the ramping of the center RF frequency. (b) Schematic of our novel acousto-optical lens (AOL), based on a double-pass configuration using crossed AODs. The laser beam is first directed to a beam splitter (BS) and then double-passes through the crossed AODs. An optional beam-shrinking telescope with magnification $M_S<1$ is used to match the beam size between the AOD aperture and the final focusing lens (e.g., a microscope objective) with focal length \(F_M\). The total focal shift \(\delta\) follows the relation given in Eq.~\ref{formula: focal shift double pass}. This shift can be enhanced by selecting AODs with larger apertures, which require a lower magnification in the beam-shrinker and thereby lead to a larger effective focal shift. (c) The double-pass configuration cancels the transverse movement automatically and ensures phase matching of the AOD. $ \textbf{k}_1$, $ \textbf{k}_2$ and $ \textbf{k}_3$  are the wave numbers of the incident, first-order deflected and retroreflected beam respectively. $ \textbf{k}_4$ is the output beam. $ \textbf{K}$ is the wave number of the sound wave. (d) Phase matching diagram for the AOD shows how the tangential phase matching automatically works for the retroreflected beam. The transparent red arrow close to $ \textbf{k}_2$ ($ \textbf{k}_3$) also shows the wide range of phase match angle which allow for large range of radio frequency.}
\label{Fig:lensing effect}
\end{figure*}

Acousto‑optic devices can rapidly modulate a laser beam’s frequency and propagation direction, and are therefore widely used. Among them, the AOD is especially attractive for wide‑range beam steering with nearly flat diffraction efficiency over its bandwidth, making it ideal for laser scanning. The output angle is determined by the RF drive frequency applied to the AOD crystal. When this frequency is swept during a scan, a chirped acoustic wave forms inside the crystal (Fig. \ref{Fig:lensing effect}(a)). For example, in an AA Opto‑electronics DTSXY‑400 AOD with acoustic velocity $v_A = 650$ m/s and a laser beam waist diameter of 4 mm, the acoustic transit time $\tau$ across the beam is $6.2~\mu\text{s}$. If the RF is ramped at $df_{\rm RF}/dt = 1~\text{MHz}/\mu\text{s}$, the RF frequency differs by $6.2~\text{MHz}$ between the leading and trailing edges of the beam. This spatial frequency gradient produces a corresponding spread in diffraction angles across the beam, i.e. a position‑dependent tilt, which acts as an effective cylindrical lens with focal length of
\begin{equation}
F_{\rm AOL}=\frac{v_A^2}{\lambda} \left(\frac{df_{\rm RF}}{dt}\right)^{-1},
\label{effective focal length}
\end{equation}
assuming that the first diffraction order is used. Here, $v_A$ is the acoustic wave speed in the crystal, $\lambda$ is the vacuum wavelength of the laser beam and $f_{\rm RF}$ is the RF frequency of the acoustic wave. As a result, AOD‑based beam steering unintentionally shifts the beam focus: the chirp‑induced spatial variation of the diffraction angle acts as a cylindrical lens that perturbs the focal position.

In neutral atom tweezer experiments, a microscope objective (possibly preceded by relay optics) is placed after the AOD to focus the beam onto the target. The transverse displacement of the focal spot is
\begin{equation}
\Delta = F_M \,\Delta_{\theta}
       = F_M \,\frac{\lambda}{v_A}\bigl(f_{\rm RF}-f_0\bigr),
\label{formula: transverse move single pass}
\end{equation}
where $F_M$ is the effective focal length of the microscope objective, $\Delta_{\theta}$ is the AOD‑induced change in beam angle, and $f_{\rm RF}$ ($f_0$) is the current (reference) RF frequency (see Fig. \ref{Fig:lensing effect}(a)). When the RF frequency is chirped, the spatial frequency gradient across the aperture makes the AOD act as an effective cylindrical lens, shifting the focus away from the original plane. Typically, the corresponding acousto‑optic lens has an effective focal length much longer than that of the objective, yet still large enough to perturb the focus during rapid scans. Approximately, the focal position change $\delta_0$ is proportional to the ramping rate of the RF frequency
\begin{equation}
\delta_0\approx-\frac{F_M^2}{F_{\rm AOL}}=-\frac{\lambda F_M^2}{v_A^2}\frac{df_{\rm RF}}{dt}.
\label{formula: focal shift single pass}
\end{equation}
The chirp‑induced effective lensing is one‑dimensional, acting only along the direction parallel to the acoustic wave. Consequently, the beam becomes astigmatic, with the two principal focal points separated by an astigmatism distance $\delta_{0}$. 

Achieving independent transverse control in both axes requires a 2D crossed‑AOD arrangement. To decouple transverse motion from longitudinal focusing, at least two pairs of 2D crossed AODs (total 4 AODs) are needed so that their transverse movement can cancel while their chirp‑induced lensing is independently controlled. In total, this four‑AOD architecture enables full three‑dimensional control of the focus position \cite{reddy_fast_2005,kirkby_compact_2010}. The concept of using the lensing effect to create an acousto-optic lens (AOL) was first introduced by Kaplan \cite{kaplan_acousto-optic_2001}. This idea was further developed by Reddy and Saggau \cite{reddy_fast_2005}, who employed a similar setup with two AODs with a 4$f$ system, enabling scanning of both focal and transverse positions. They also highlighted that with two pairs of AODs, 3D scanning becomes feasible. This was subsequently implemented by the same group \cite{duemani_reddy_three-dimensional_2008} in a two-photon excitation microscope for the imaging of biological materials.

An adaptation of this system involved the use of four cascaded AODs to form a compact AOL \cite{kirkby_compact_2010}. Typically, AODs are designed with birefringent materials, which align two wave vectors with the acoustic wave vector under a tangential condition. This expands the output frequency range, but restricts the incident angle range to approximately $\pm$1 mrad \cite{kirkby_compact_2010}. There is an issue with the incident angle changing due to the varying angle at which the beam from the first AOD enters the second. By employing specialized, optically and acoustically rotated AODs, this limitation was overcome, albeit with a constrained output beam range \cite{kirkby_compact_2010}. Since then, this system has gained popularity in the TPEF community, as evidenced by various studies \cite{katona_fast_2012,nadella_random-access_2016}.

In addition, a similar idea of using the acousto-optic effect to change the focus of a laser beam has been proposed \cite{mermillod-blondin_high-speed_2008}, called the tunable acoustic gradient index of refraction lens (TAG lens). Using an inhomogeneous refractive index of a material in a barrel, the device provides good optical power similar to that of the aforementioned AOLs. However, the TAG lens only supports resonant scanning, rather than raster scanning \cite{mermillod-blondin_high-speed_2008}, which limits its use in tweezer array experiments. More review articles on tunable lenses using various methods can be found here \cite{kang_variable_2020,chen_electrically_2021}.

\section{Design of a double-pass type AOL}\label{sec: AOL design}

We designed a modified version of the AOL using a double-pass configuration to self-cancel the transverse movement during RF chirping, inspired by the conventional double-pass structure for using an AOM with a wider RF frequency range without changing the beam direction \cite{donley_double-pass_2005}. The laser beam is injected into the cross-aligned AODs and then focused and retro-reflected by a cat-eye structure, as shown in Fig. \ref{Fig:AOL}(a). This functional block acts purely as a focal shifter, without affecting the transverse position of the beam. To achieve transverse focus movement, the beam is subsequently sent through another crossed AOD pair. Together, this configuration forms an AOL capable of fully 3D atom shuttling in optical tweezers.

Note that the AOD operates in shear mode. As depicted in Fig. \ref{Fig:lensing effect} (c) and (d), the k-vector plot shows the phase matching of a birefringent material of an AOD. The incident laser beam is marked as \( \textbf{k}_1\), which lies on the e-elliptical sphere. The single-passed and retroreflected laser beams, marked as \( \textbf{k}_2\) and \( \textbf{k}_3\), lie on the o-elliptical sphere. The double-passed laser beam, marked as \( \textbf{k}_4\), also lies on the e-elliptical sphere. The acoustic wavevector is marked with \( \textbf{K}\), which needs to be phase matched for the input and output beams. The polarization of the laser beam from the e-elliptical sphere is perpendicular to that on the o-elliptical sphere. In our case, with crossed AODs, the polarization is rotated twice for a single pass, which means that we have the same polarization of the beam everywhere except between the two AODs. As a result, after a double pass, the beam retains its original polarization, which necessitates the use of an optical circulator to redirect it, as shown in Fig. \ref{Fig:AOL}(a).

To test such a setup, as shown in Fig. \ref{Fig:lensing effect}(b), we simplify the configuration by replacing the optical circulator with a beam splitter (BS). Although this introduces power loss, it does not affect the measurement results. We used commercial AA Optics AODs (DTSXY-400) and an input beam with a 2$\sigma$ waist of 4 mm. Each AOD is controlled by an arbitrary wave generator (AWG) with an RF center frequency of 100 MHz. The first order diffraction of both AODs are selected by an aperture and then retroreflected by a cat-eye with a lens of \(F=100\) mm and a mirror. After double passing, we add a final lens $F_M=100$ mm for the final focus and for measuring the focal shift \(\delta\). Note that an optional beam-shrinking telescope with magnification $M_S<1$ can be used to match the beam size between the AOD aperture and the final focusing lens. The telescope can increase the tuning range of the focal shift, however we did not add it in our test setup. In the remainder of the text $M_S=1$ if it is not otherwise mentioned.

\begin{figure*}[htbp]
\centering
\includegraphics[width=0.8\linewidth]{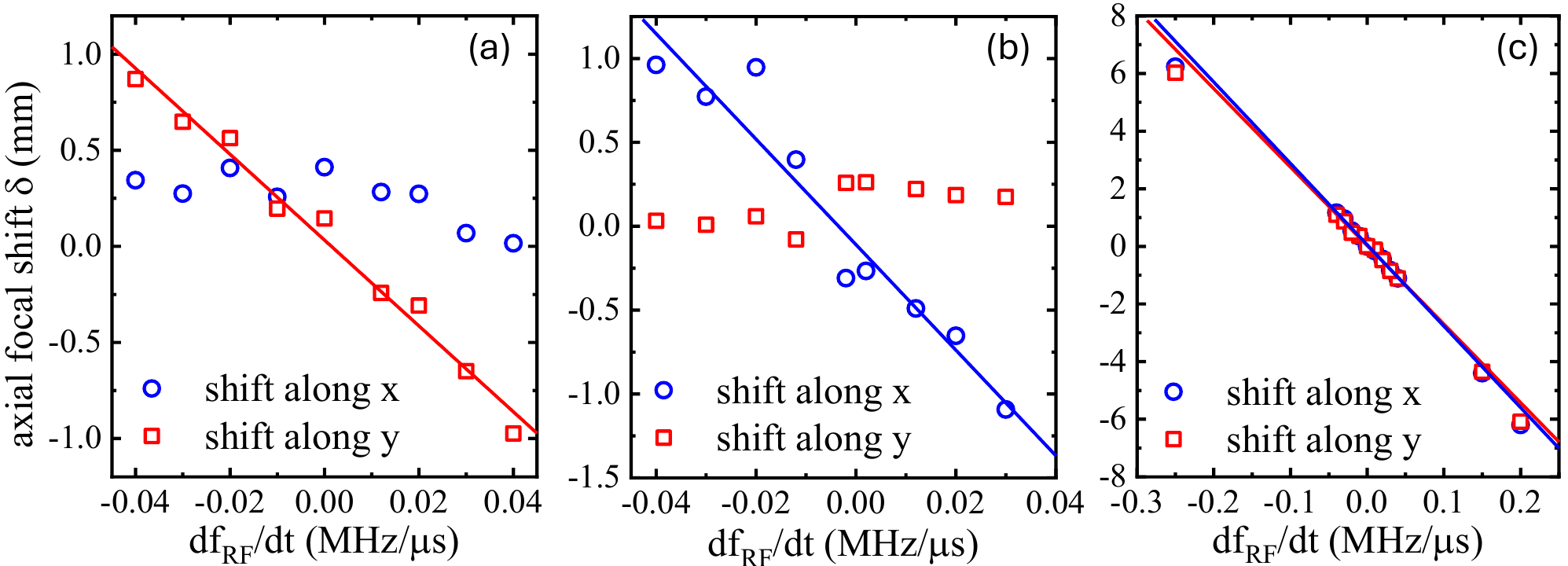}
\caption{The AOL is characterized at static focus positions, where each data point corresponds to a fixed focal position during measurement. (a) and (b) illustrate the change in focus when only one AOD is driven with a chirped RF signal, which manifests as astigmatism. Here, \(df_{\rm RF}/dt\) denotes the RF frequency ramping rate, which can be positive or negative and thereby induces the corresponding focal shift \(\delta\). The slope of the resulting curve yields the parameter \(-2\lambda F_M^2/v_A^2\), with \(F_M = 100~\text{mm}\) in this test setup. (c) shows that when both AODs are driven with identical linearly chirped RF signals, a focal shift is observed along both \(x\)- and \(y\)-directions, effectively making the system a spherical lens.}
\label{Fig: static test}
\end{figure*}

For a single pass through the AOD, the chirping RF signal forms a cylindrical lens with an effective focal length, as described in Eq. (\ref{formula: focal shift single pass}). After double passing, we achieve an additional effective lens, resulting in a final focal shift of
\begin{equation}
\delta\approx-\frac{2F_M^2}{F_{\rm AOL}M_S^2}=-\frac{2\lambda F_M^2 }{v_A^2M_S^2} \frac{df_{\rm RF}}{dt}.
\label{formula: focal shift double pass}
\end{equation}
The range of the focal shift \(\delta\) is determined by the final focus lens $F_M$, magnification of the beam shrinker $M_S$, and the ramping speed of the RF signal. Note that by selecting an AOD with a larger aperture, e.g. Gooch \& Housego AODF 4085, a larger demagnification is required, and thereby the achievable focal shift increases by a factor of \(M_S^{-2}\). The maximum RF ramp speed is another limit which is related to the beam waist and the acoustic bandwidth \(f_{\rm span}\) of the AOD. The shortest possible chirp duration is approximately \(\tau\), the time it takes the RF signal to traverse the optical beam, leading to a maximum chirp rate of \(f_{\rm span}/\tau\). However, if the RF frequency needs to be sustained for a certain duration \(T\), the effective ramp speed must be reduced to \(f_{\rm span}/(\tau + T)\). Moreover, the finite sound velocity also manifests itself as a temporal delay between the applied RF signal and the resulting optical focal shift: the laser interacts with the preloaded acoustic wave, which requires time to propagate across the crystal.

\section{Charaterization of the performace of the double passed AOL}\label{sec: characterization}

To validate the functionality of the acousto-optical setup, we performed two types of optical tests: the static-focus test and the dynamic-focus test. In the static-focus test, we linearly ramped the RF frequency to investigate the shift of the static focal position. A sawtooth waveform with varying ramp rates was applied, and the resulting laser beam was monitored using a beam profiler. In the dynamic-focus test, we aimed to simulate the conditions of a future tweezer experiment by employing a nonlinear RF chirp to dynamically steer the focal point. This was achieved by repetitively playing back a predefined RF pulse sequence into the AOD while pulsing the laser beam using an external trigger with adjustable delay relative to the RF signal. This approach allowed us to capture the instantaneous beam profile at a given focus position.

\begin{figure*}[htbp]
\centering
\includegraphics[width=0.9\linewidth]{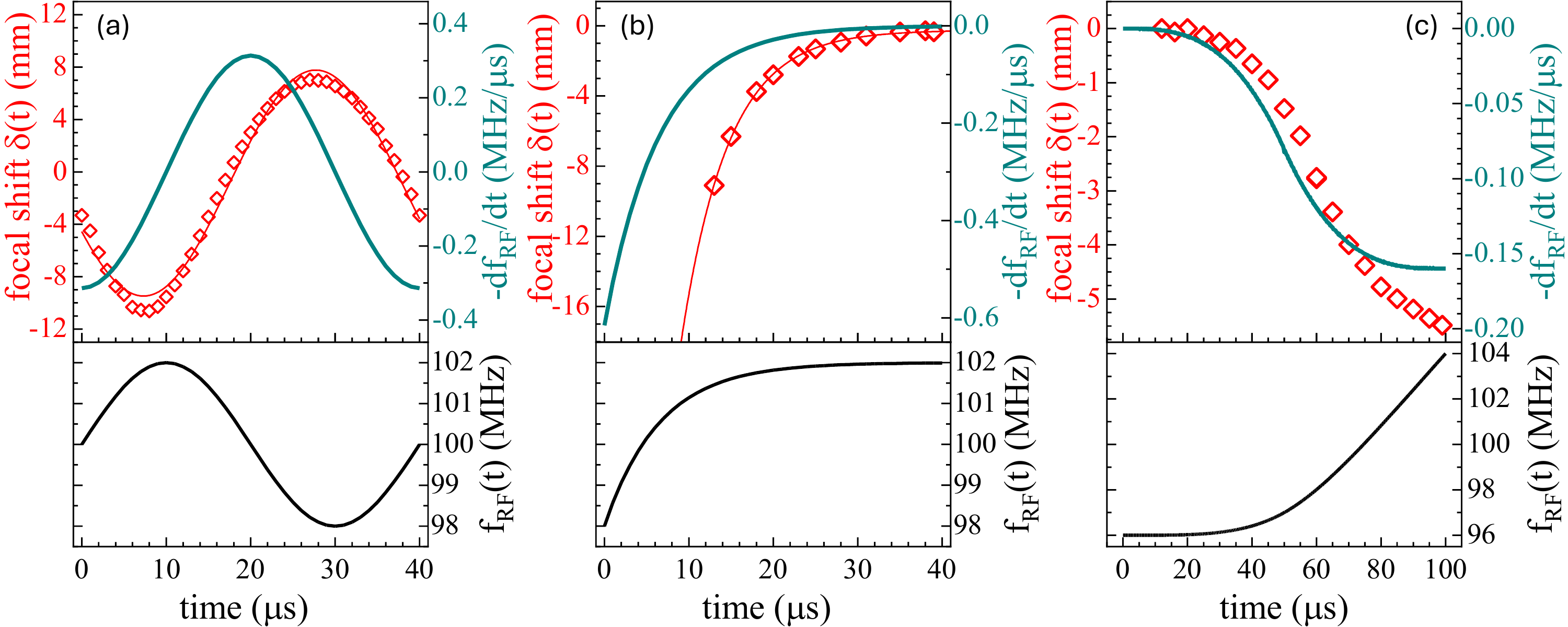}
\caption{The AOL is characterized using arbitrary RF frequency trajectories in a dynamical focus test. Figures (a)–(c) present measurements for three different RF frequency profiles: sinusoidal, exponential decay, and constant-jerk. The bottom panels display the applied RF frequency trajectories (black line) $f_{\rm RF}(t)$, while the top panels show the negative derivatives of RF frequency trajectories (green line) $-df_{\rm RF}/dt$ and the corresponding measured focal-position shifts (red diamond) $\delta(t)$, where the vertical axes of them are linked by the parameter \(2\lambda F_m^2 / v_A^2\), whose value (27.8 mm/(MHz/$\mu$s)in this setup) is obtained from average of two directions in the static-focus measurements. The measured focal shifts are fitted with the target trajectory functions (red line). The observed time delay between RF ramping speed and focal position and focal-shift distortions are discussed in Sec.~\ref{sec: characterization}.
}
\label{Fig: dynamical test}
\end{figure*}

In the static-focus test, a linearly chirped RF frequency is applied to one (or both) AOD(s), which functions as a tunable cylindrical (or spherical) lens. When only one AOD is chirped, as illustrated in Fig. ~\ref{Fig: static test} (a) and (b), the acousto-optical lens (AOL) acts as a cylindrical lens, shifting the focal position only along the axis corresponding to the modulated AOD. A positive (negative) frequency ramp corresponds to a convex (concave) lensing behavior. The results clearly exhibit astigmatism: the focus positions of $x$ or $y$ are separated. By linearly chirping both AODs with identical waveforms, we achieve focal shifts in both directions, effectively turning the crossed AOD configuration into a tunable spherical lens, as shown in Fig. ~\ref{Fig: dynamical test}(c). The linear fit to the data gives a slope of -28.3(-27.3) mm/(MHz/$\mu$s) in the $x$($y$) direction, which is close to the theoretical value \(-2\lambda F_m^2 / v_A^2\)=-33.0 mm/(MHz/$\mu$s). The discrepancy arises from a slight misalignment of the cat-eye retroreflector, which deviates from the ideal 4$f$ configuration, resulting in an additional effective magnification factor and modifying the overall focal shift.

For the dynamic-focus test, the main difference is the introduction of a fiber-coupled AOM before the laser beam enters the beam splitter (as shown in Fig. \ref{Fig:lensing effect}(b)). The fiber AOM is triggered by the RF output from the AWG, allowing the laser to be pulsed at various time delays relative to the chirped RF waveform applied to the AODs. The laser pulse duration is about 50 ns which is sufficiently short for the camera to resolve the instantaneous focal position. The AWG continuously outputs a chirped RF waveform. At each chosen time delay, the fiber AOM is triggered to generate a laser pulse, which is monitored by a photodiode after the transmission through the beam splitter to ensure synchronization. Finally, the spatial profile of the focused beam is recorded by a camera positioned after the final lens \(F_M\).

By plotting the focus shift \(\delta(t)\) as a function of the delay time, we demonstrate that arbitrary focus trajectories can be engineered, for example, sinusoidal modulation, exponential decay, and constant-jerk motion, as shown in Fig.~\ref{Fig: dynamical test}(a-c). In these experiments, both AODs are driven by identical chirped RF waveforms, effectively forming a spherical lens. The control RF frequency profile \(f_{\rm RF}(t)\) corresponding to a given focus trajectory \(\delta(t)\) is obtained by integrating the RF ramping speed related to Eq. ~\ref{formula: focal shift double pass}. In the bottom panels of Fig. ~\ref{Fig: dynamical test}, the RF frequency trajectory \(f_{\rm RF}(t)\) is plotted. Its negative derivative \(-df_{\rm RF}/dt\) is then shown together with the measured focal shift $\delta(t)$ in the upper panels for comparison. Note that the vertical axes of the focal shift and the RF ramping speed are linked by a calibration coefficient obtained from the static-focus measurement (average of two transverse directions). The control trajectory consistently leads the focal-shift trajectory by $\sim$8~\(\mu\)s. This offset arises because the AOD introduces an intrinsic delay due to the finite speed of sound propagation in the crystal, as discussed in Sec. ~\ref{sec: AOL design}. Overall, the agreement between theory and experiment is good across all cases, with only minor deviations observed when the second derivative of the trajectory becomes large. These deviations are attributed to nonlinear acoustic effects during large chirp rates, which distort the effective lensing profile.

\section{Towards improvement of atom shuttling in tweezer array experiment}\label{sec: applications}

With this powerful tool, we can now consider the benefits it provides for atom shuttling in a neutral atom quantum computer. We discuss two scenarios in this paper: the first is to compensate for the detrimental lensing effect during the in-plane coherent transportation by AODs; the second is shuttling atoms out of the plane to facilitate processes such as rearrangement, fully connected entanglement gates and local imaging.

\subsection{Tweezer trap deformation due to the lensing effect of AODs}\label{sec:lensing issue}

Coherent atom transport is essential for achieving full connectivity in neutral atom quantum computers. A key application is error correction of logical qubits, where the entanglement of many spatially separated atoms is required. This is particularly relevant for recently proposed codes, such as the LDPC code \cite{pecorari_high-rate_2025,xu_constant-overhead_2024}, which require entanglement between large amounts of distant qubits. Thus, maintaining both high survival rates and quantum coherence during transport is critical to the fidelity of logical gate operations. Moreover, the duration of the transport process directly affects the overall computation speed and constrains the allowable circuit depth. As analyzed in several studies \cite{wang_efficient_2024}, the fidelity and efficiency of atom transport are crucial for enabling practical large-scale quantum error correction.

\begin{figure*}[htbp]
\centering
\includegraphics[width=0.9\linewidth]{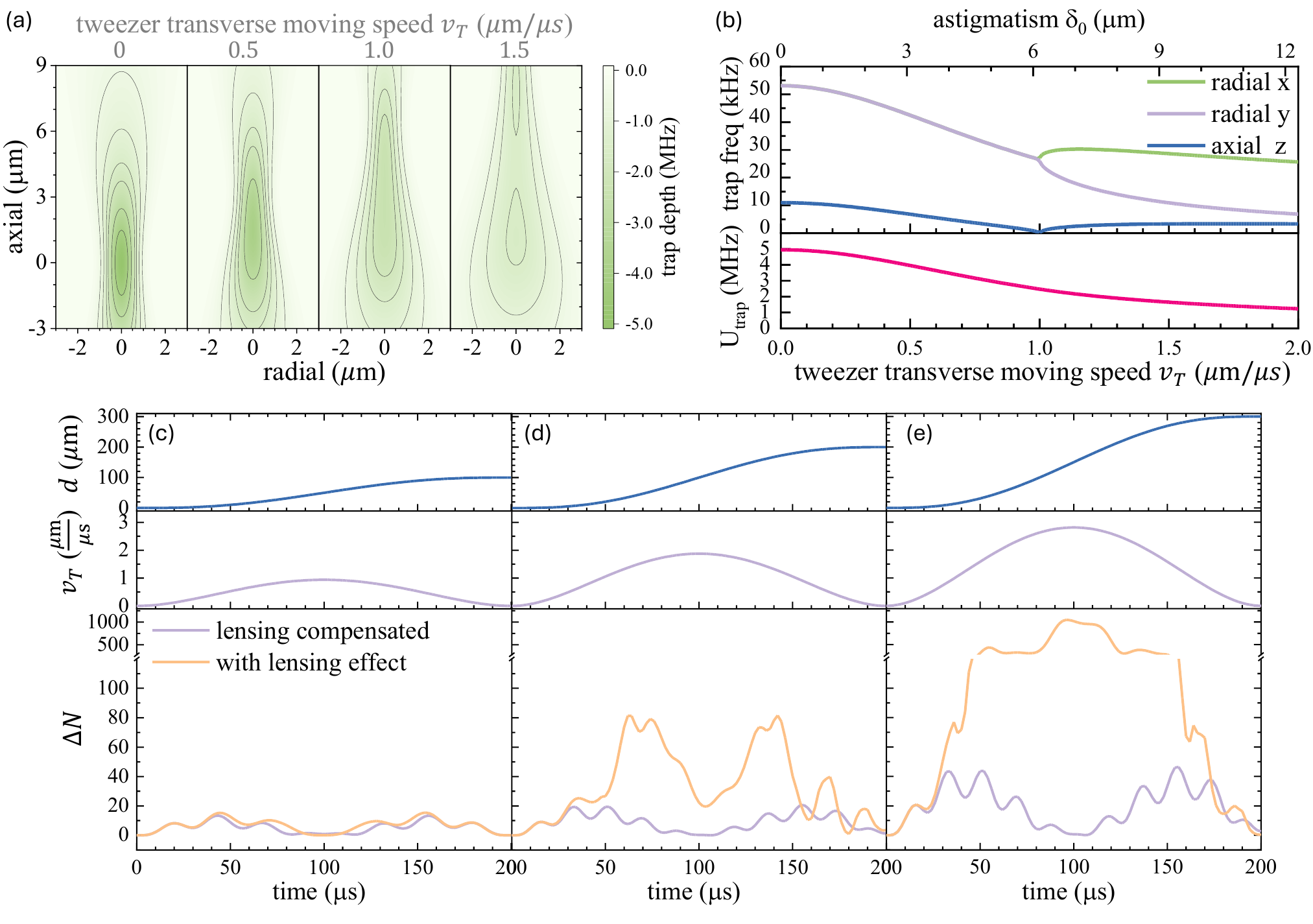}
\caption{
Simulation of tweezer trap deformation and atomic motional excitation due to the lensing effect of a single AOD during transverse motion. The simulation is based on a 0.5~mW, 828~nm laser focused by a microscope objective with a focal length of 4~mm, producing a tweezer with a waist of 0.9~\(\mu\)m. The AOD used is the DTSXY-400 from AA Optics, with an acoustic velocity of 650~m/s.  
(a) Simulated tweezer potentials at various transverse speeds \(v_T\), showing trap deformation, reduction in trap depth, and eventual splitting into two potential wells when \(v_T > 1~\mu\text{m}/\mu\text{s}\).  
(b) Trap depth (pink) $U_{\rm trap}$ and trap frequencies (radial \(x, y\) and axial \(z\)) as functions of transverse speed. The axial frequency drops to zero around \(v_T = 1.0~\mu\text{m}/\mu\text{s}\), indicating the onset of trap splitting. Note that the top axes shows the astigmatism (focal shift) $\delta_0$ which is proportional to the transverse moving speed of the tweezer by Eq. \ref{astig vs speed}.
(c) - (e) Simulations of atom transport over 200~\(\mu\)s using minimum-jerk trajectories for various transport distances. Longer distances result in higher peak velocities, which enhance the lensing effect and increase motional excitation. The bottom panels show the time evolution of the average motional excitation \(\Delta N\), both with and without lensing compensation. When \(v_T\) exceeds 1.0~\(\mu\text{m}/\mu\text{s}\), excitation grows rapidly in the uncompensated case. Note that the ratio of trap depth to trap frequency sets a threshold for atom loss. With our proposed AOL-based compensation scheme, the trap deformation can be dynamically corrected, leading to significantly reduced excitation and enabling faster, high-fidelity atomic transport.
}
\label{Fig: lensing of single AOD}
\end{figure*}

As discussed in Sec.~\ref{sec: AOL design}, using an AOD to shuttle atoms in-plane introduces a lensing effect that leads to astigmatism \(\delta_0\) in the tweezer beam. This astigmatism is proportional to the transverse velocity of the trap. In Fig. ~\ref{Fig: lensing of single AOD}, we simulate the case of focusing a 0.5 mW of 828~nm laser using a microscope objective with focal length \(F_M = 4~\text{mm}\), yielding a final beam waist of 0.9~\(\mu\)m for trapping a \({}^{87}\)Rb atom. Figure~\ref{Fig: lensing of single AOD}(a) illustrates how the tweezer shape deforms at different transverse speeds $v_T$. At lower speeds, below 1.0~\(\mu\text{m}/\mu\text{s}\), the trap undergoes deformation, leading to a reduction in both trap depth and trap frequencies, as shown in Fig.~\ref{Fig: lensing of single AOD}(b). Notably, when the speed exceeds 1.0~\(\mu\text{m}/\mu\text{s}\), the trap splits into two distinct potential wells. This transition is also marked by the axial trap frequency dropping to zero as shown in Fig.~\ref{Fig: lensing of single AOD}(b).

\begin{figure*}[htbp]
\centering
\includegraphics[width=0.9\linewidth]{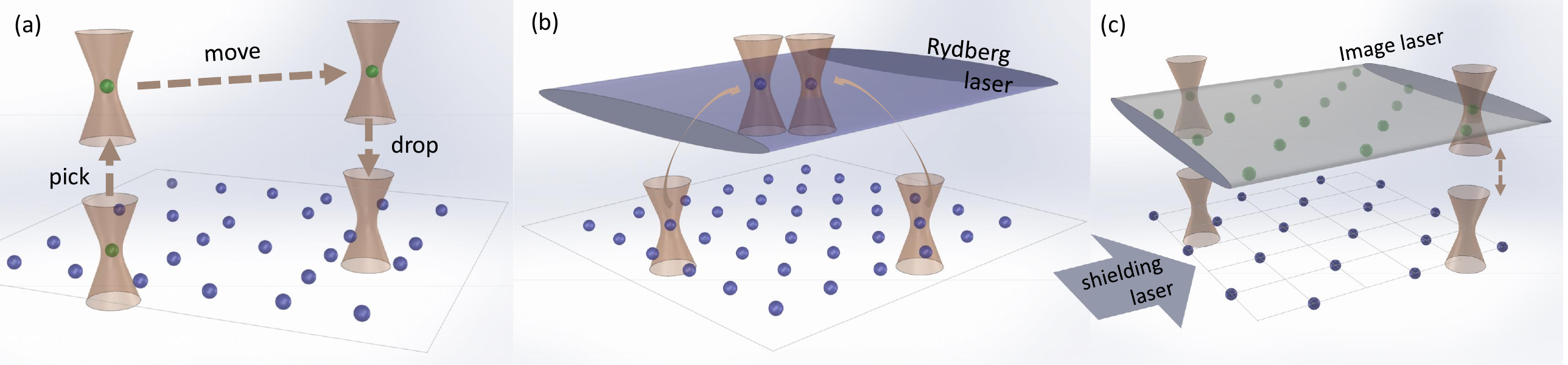}
\caption{Applications of off-plane shuttling in a tweezer array. (a) An atom can be lifted out of the static tweezer plane and transported to a target location without colliding with other atoms. This method also enables parallel shuttling of multiple atoms. (b) Two or more atoms can be raised to a plane containing a Rydberg laser to perform entangling operations, thereby enabling long-range entanglement and ensuring full connectivity across the array. (c) Atoms can be moved to a separate imaging plane for near-in-situ fluorescence detection, facilitating mid-circuit measurement. To protect atoms remaining in the original plane from scattering, a shielding laser is applied to the original plane.}
\label{Fig:applications}
\end{figure*}

When transporting atoms with the aforementioned lensing effect, low transverse speeds result in reduced trap depth and weakened trap frequencies, making atoms more susceptible to motional excitation. This is because the excitation probability is inversely proportional to the trap frequency and is also related to the Fourier transform of the acceleration profile at the trap frequency \cite{carruthers_coherent_1965}. In Fig.~\ref{Fig: lensing of single AOD}(c)–(e), we present a simulation of atom transport over 200~\(\mu\)s with varying transport distances. All cases utilize the minimum-jerk trajectory, which performs well for long-distance shuttling \cite{liu_molecular_2019,cicali_fast_2025,hwang_fast_2024}. As the transport distance increases, the average and peak tweezer velocities also increase, where the lensing effect leads to stronger atomic excitation. We simulate the average motional excitation level \(\Delta N\) as a function of time (for details see Supplementary Material, Sec. \ref{sec: appendix}), shown in the bottom panels of Fig. \ref{Fig: lensing of single AOD}. When the average transport speed exceeds 1~\(\mu\text{m}/\mu\text{s}\), a significant increase in excitation is observed due to trap deformation. To mitigate this, we can compensate by using the AOL described earlier. By dynamically adjusting the AOL's focal position in the opposite direction of the lensing distortion, we restore the trap shape during transport. For instance, when transporting an atom along the transverse \(x\)-direction, we can simultaneously ramp the RF frequency in the \(x\)-channel of the AOL to shift the focus oppositely, effectively canceling the deformation.

\subsection{Off-plane shuttling for rearrangement, two-qubit gate and imaging}\label{sec: off-plane shuttlling}

Generating a defect-free atom array is a critical requirement for neutral-atom quantum computing. The standard approach involves stochastic loading followed by fluorescence imaging to determine atom positions, after which atoms are rearranged into the target configuration using movable tweezers. However, conventional methods are restricted to in-plane motion, and avoiding collisions during rearrangement introduces additional algorithmic complexity. Here, we propose using the AOL to perform out-of-plane atom shuttling, as illustrated in Fig.~\ref{Fig:applications}(a). In our scheme, atoms are first lifted out of the static tweezer plane, then transported to their target positions, and finally lowered back into the array. Given the limited duration with which the AOL can sustain certain RF chirping rates, we define the upper plane as the zero-chirp baseline. This allows atoms to be held out of plane for extended periods while transverse movement can be performed, enabling flexible and collision-free rearrangement.

Atom shuttling also plays a key role in enabling entanglement between distant qubits. Rather than moving atoms laterally to a separate interaction region \cite{bluvstein_logical_2023, chiu_continuous_2025}, we propose an alternative approach in which the Rydberg interaction zone is positioned directly above the original tweezer array, as illustrated in Fig.~\ref{Fig:applications}(b). In this scheme, two or more atoms can be lifted out of the plane, effectively jumping over the rest of the array, and moved to a plane illuminated by a Rydberg excitation beam, where an entangling gate can be performed. This approach can be especially useful in quantum error correction: for instance, by driving multiple RF tones on the transverse AODs, one can simultaneously lift an entire logical qubit array, shuttle it to the interaction layer, perform multi-qubit Rydberg gates, and then return the atoms to their original positions. This technique enables fast, scalable, and collision-free entanglement operations critical for fault-tolerant quantum computation.

The same out-of-plane shuttling technique can also be applied to perform imaging in a separate plane, as illustrated in Fig.~\ref{Fig:applications}(c). The primary challenge in such a scheme is the limited vertical separation between the imaging and storage planes, which can result in scattered fluorescence photons being reabsorbed by atoms remaining in the storage layer. This issue can be mitigated by applying a shielding laser to the storage plane to protect the atoms from stray fluorescence during imaging. Similarly to the two-qubit gate protocol, atoms can be selectively lifted for imaging, enabling partial readout of the array without disturbing the rest. Since only a modest vertical displacement of 15–20~\(\mu\)m is required, this approach significantly reduces the mechanical transfer distance and enables fast, localized imaging without crosstalk, thereby enhancing the overall measurement speed and fidelity.

\section{Conclusion and outlook}

In this paper, we introduced novel modifications to the acousto-optical lens (AOL) for neutral atom quantum computing, enhancing the ability to shuttle atoms in three-dimensional space. Our approach addresses the limitations of existing crossed-AOD-based methods, which restrict atom movement to a single plane. By incorporating a double-pass AOD configuration for focal tuning alongside normal crossed AODs for transverse tweezer movement, we can improve atom rearrangement with tweezers, overcoming speed limitations due to the lensing effect of AODs. Our optical test results demonstrated the new AOL's capabilities in achieving arbitrary focus position movements, which opens new possibilities for advanced 3D atom shuttling in neutral atom quantum computers. Moreover, the same approach can be used in cold‑atom experiments to "paint" 3D optical dipole trap geometries, extending the conventional 2D painting method \cite{henderson_experimental_2009}.

\section*{ACKNOWLEDGEMENTS}
The authors thank M. Venderbosch, M. Festenstein, R. Lous, J. del Pozo Mellado, D. Janse van Rensburg, R. Venderbosch, R. Teunissen, and Y. van der Werf of the Sr and Rb quantum computing teams at Eindhoven University of Technology (https://www.tue.nl/rydbergQC) and J. He, I. Knottnerus, A. Urech, Y. Tseng, R. Spreeuw, and F. Schreck of the SrPAL and SrMic teams at the University of Amsterdam (https://www.strontiumbec.com) for their support and collaboration. This research is financially supported by the Dutch Ministry of Economic Affairs and Climate Policy (EZK), as part of the Quantum Delta NL program, the Horizon Europe programme HORIZON-CL4-2021-DIGITAL-EMERGING-01-30 via the project 101070144 (EuRyQa), and by the Netherlands Organisation for Scientific Research (NWO) under Grant No.\ 680.92.18.05. 

\bibliography{main}

\newpage
\section{Supplementary Material}
\label{sec: appendix}

To quantitatively assess the impact of the AOD lensing effect on atom transport, we evaluate two key aspects:
(1) the deformation of the optical tweezer due to the time-varying focal shift induced by the AOD;  
(2) the resulting motional excitation of a trapped atom during transport under such a distorted potential.

\subsection*{Trap deformation induced by AOD lensing}

We consider an optical tweezer formed by focusing a 0.5~mW, 828~nm laser beam using a high NA microscope objective with a focal length of \(F_M = 4~\text{mm}\), producing a beam waist of \(w_0 = 0.9~\mu\text{m}\) at the focus. This tweezer is used to trap a single \({}^{87}\)Rb atom. 

Due to the chirped RF signal applied to the AOD, a spatially varying diffraction angle is introduced, effectively forming a cylindrical lens that shifts the beam waist differently in the $x$ and $y$ directions. This leads to asymmetric beam waists during transport:

\begin{align}
w_x &= w_0 \sqrt{\left(\frac{\lambda z}{\pi w_0^2}\right)^2 + 1} \\
w_y(\delta_0) &= w_0 \sqrt{\left(\frac{\lambda (z - \delta_0)}{\pi w_0^2}\right)^2 + 1}
\end{align}

where $\delta_0$ represents the effective focal shift in the $y$ direction induced by the AOD chirp. It is proportional to the ramping speed of the RF frequency $df_{\rm RF}/dt$:

\begin{align}
\delta_0 = -\frac{\lambda F_M^2}{v_A^2} \frac{df_{\rm RF}}{dt}
\end{align}

On the other hand, the RF ramping speed is also related to the transverse moving speed of the tweezer:

\begin{align}
\frac{df_{\rm RF}}{dt} = \frac{v_A}{F_M \lambda} v_T
\end{align}

where $v_T$ is the transport speed of the tweezer focus. Then we have a relation between the focal shift and the transverse moving speed:

\begin{align}
\label{astig vs speed}
\delta_0 = -\frac{\lambda F_M}{v_A \lambda} v_T
\end{align}

which are also shown as the top and bottom axes in Fig. \ref{Fig: lensing of single AOD}(b). The resulting intensity profile of the asymmetric Gaussian beam becomes:

\begin{equation}
I(x, y, \delta_0) = \frac{2P}{\pi w_x w_y(\delta_0)} \exp\left(-\frac{2x^2}{w_x^2} - \frac{2y^2}{w_y(\delta_0)^2}\right)
\end{equation}

The corresponding optical trapping potential, including gravity, is as follows:

\begin{equation}
\begin{aligned}
U(x, y, z, &\delta_0) = -\frac{3 \pi c^2 \Gamma I(x,y,\delta_0)}{2 \omega^3_{\rm trap}} \times\\
&\left( \frac{1}{\omega_{\rm atom} + \omega_{\rm trap}} + \frac{1}{\omega_{\rm atom} - \omega_{\rm trap}} \right) 
- m g z
\end{aligned}
\end{equation}

As the trap asymmetry evolves during transport, both the depth and trapping frequencies are distorted, which leads to time-dependent variations in confinement and potential heating, as shown in Fig. \ref{Fig: lensing of single AOD}(a) and (b).

\subsection*{Motional excitation under time-dependent potential}

We now analyze the motional excitation of an atom under the time-dependent trap described above. The effective one-dimensional Hamiltonian in dimensionless units is given by:

\begin{equation}
\label{hamiltonian}
H(t) = \frac{p^2}{2} + \frac{1}{2}\omega(t)^2x^2  - g(t)x
\end{equation}

where, $\omega(t)$ is the dimensionless time-varying trapping frequency ($\omega(t=0)=1$ as initial condition), and $-g(t)x$ is an effective inertial force arising from the tweezer acceleration. Note that here we simplify the potential to be harmonic. More precise treatment can be found in Ref. \cite{cicali_fast_2025,hwang_fast_2024}. Assuming the atom is initially in the ground state of the harmonic trap:

\begin{equation}
\psi(x, t=0) = \frac{1}{\pi^{1/4}} e^{-x^2/2}
\end{equation}

Note that the state remains a displaced coherent state during evolution\cite{carruthers_coherent_1965-1,sturzu_explicit_2001}, leading to the following form:

\begin{equation}
\psi(x,t) \sim \exp \left( \sum_{n=0}^2 a_n(t) x^n \right)
\end{equation}

with initial conditions:
\begin{align}
a_0(0) &= a_1(0) = 0 \\
a_2(0) &= -\frac{1}{2}
\end{align}

The time evolution of the coefficients is governed by the following set of coupled differential equations:

\begin{align}
\frac{d a_0(t)}{dt} - \frac{i}{2} a_1(t)^2 - ia_2(t) &= 0 \\
\frac{d a_1(t)}{dt} - 2ia_1(t) a_2(t) - i g(t) &= 0 \\
\frac{d a_2(t)}{dt} - 2i a_2(t)^2 + \frac{i}{2} \omega(t)^2 &= 0
\end{align}

From the evolved wavefunction, we can extract the average vibrational quantum number $\Delta N$ as:

\begin{equation}
\Delta N = \frac{\omega(t)\left[ \Re(a_1(t))^2 - a_2(t) \right]}{8 a_2(t)^2} + \frac{\Im(a_1(t))^2 - a_2(t)}{2 \omega(t)} - \frac{1}{2}
\end{equation}

which is also plotted in Fig. \ref{Fig: lensing of single AOD}(c). This simplified calculation characterizes the motional excitation induced during the tweezer transport.

\subsection*{Minimum-jerk transport and motional state excitation}

To minimize motional heating, a minimum-jerk trajectory \cite{liu_molecular_2019,cicali_fast_2025,hwang_fast_2024} is often used for atom transport:

\begin{align}
x(t) &= d\left(6 \frac{t^5}{T^5} - 15 \frac{t^4}{T^4} + 10 \frac{t^3}{T^3} \right) \\
v(t) &= d\left(30 \frac{t^4}{T^5} - 60 \frac{t^3}{T^4} + 30 \frac{t^2}{T^3} \right) \\
g(t) &= d\left(120 \frac{t^3}{T^5} - 180 \frac{t^2}{T^4} + 60 \frac{t}{T^3} \right)
\end{align}

Here, $d$ is the total transport distance, $T$ is the transport duration, and $g(t)$ is the acceleration which is used in the Hamiltonian (Eq. \ref{hamiltonian}) as the inertial force term.

By numerically solving the above system with and without lensing-induced focal shift $\delta_0$, we can compare the resulting trap deformation and excitation. Simulations show that the AOD lensing effect leads to anisotropic deformation of the tweezer, reduced trap depth, and increased excitation probability during transport, indicating that compensation of such lensing is critical for achieving high-fidelity coherent atom transport, as shown in Sec. \ref{sec:lensing issue}.

\end{document}